\documentclass{PoS}

\title{Challenges in reconciling observations and theory of the brightest high-energy flare ever of 3C~279}

\ShortTitle{3C~279, the flare of June 2015}

\author{\speaker{Eugenio Bottacini}\thanks{NASA grant NNX13AO84G is acknowledged.}\\
        Dipartimento di Fisica e Astronomia "G. Galilei", Universita di Padova, I-35131 Padova, Italy\\
        W.W. Hansen Experimental Physics Laboratory \& Kavli Institute for Particle Astrophysics and Cosmology, Stanford University, USA\\
        E-mail: \email{eugenio.bottacini@unipd.it}}

\author{Markus B{\"o}ttcher\\
       Centre for Space Research, North-West University, Potchefstroom 2531, South Africa}

\author{Elena Pian\\
       INAF, Istituto di Astrofisica Spaziale e Fisica Cosmica, via P. Gobetti 101, 40129 Bologna, Italy\\
       Scuola Normale Superiore, Piazza dei Cavalieri 7, 56122 Pisa Italy\\
      INFN, Sezione di Pisa, Largo Pontecorvo 3, 56127 Pisa, Italy}
 
 \author{Werner Collmar\\
       Max-Planck-Institut f{\"u}r extraterrestrische Physik, Giessenbach, 85748 Garching, Germany}
 
 \author{Dario Gasparrini\\
        Agenzia Spaziale Italiana (ASI) Space Science Data Center, I-00133, Roma, Italy\\
        Istituto Nazionale di Fisica Nucleare, sezione di Perugia, I-06123, Perugia, Italy}

\abstract{Recent high-energy missions have allowed keeping watch over blazars in
flaring states, which provide deep insights into the engine powered by supermassive
black holes. However, having a quasar caught in a very bright flaring state is not easy
requiring long surveys. Therefore, the observation of such flaring events represents a
goldmine for theoretical studies.\\
Such a flaring event was captured by the {\em INTEGRAL} mission in June 2015 while
performing its (as of today) deepest extragalactic survey when it caught the prominent
blazar 3C~279 in its brightest flare ever recorded at gamma-ray energies. The flare
was simultaneously recorded by the {\em Fermi} gamma-ray mission, by the {\em Swift} mission, by the
{\em INTEGRAL} mission and by observations ranging from UV, through optical to the
near-IR bands. The derived snapshot of this broad spectral energy distribution of
the flare has been modeled in the context of a one-zone radiation transfer leptonic
and lepto-hadronic models constraining the single emission components. The derived
parameters of both models challenge the physical conditions in the jet. However,
very recently published very-high-energy (VHE) data at TeV energies are very close to our
lepto-hadronic model.}

\FullConference{35th International Cosmic Ray Conference --- ICRC2017\\
		10--20 July, 2017\\
		Bexco, Busan, Korea}

\begin{document}

\section{Introduction}
The flat spectrum radio quasar (FSRQ) 3C~279 has gained prominence due to its bright flaring state
detected by the Energetic Gamma Ray Experiment Telescope (EGRET) on board the
{\em Compton Gamma-Ray Observatory} (CGRO) mission \cite{hartman92}. This blazar, at redshift
$z$=0.5362 \cite{marziani96}, was the first (of currently only 5) FSRQs detected by ground-based
Atmospheric Cherenkov Telescope facilities at very-high-energy (VHE: $E > 100$~GeV) gamma rays
\cite{Albert08, mirzoyan15}. This makes it an ideal target for multifrequency studies. Such extensive
multifrequency campaigns \cite{collmar10} have allowed constraining important single emission components
from the source \cite{pian99}.\\
Here we present the results from of the multifrequency campaign that pivots around the detection
by the {\em INTEGRAL} mission of the brightest high-energy flare ever displayed by 3C~279 in
June 2015. We discuss the challenges faced by the modeling of the SED with a leptonic and
a lepto-hadronic model shown in our recent work \cite{bottacini16} and the implications of recently
analyzed TeV data.

\section{Observations}
- {\em INTEGRAL} -- In June 2015 {\em INTEGRAL} was performing its deepest extragalactic survey on the Come sky
region. Due to the huge field of view of 29 $\times$ 29 deg$^2$ the imager IBIS \cite{ubertini03}
detected the outburst of 3C~279 in only 50 ks allowing us to compute a spectrum at energies
above 15 keV. Data analysis is performed with the standard Off-line Scientific Analysis (OSA)
software\footnote{http://www.isdc.unige.ch/integral/analysis\#Software} provided by the {\em INTEGRAL}
Science Data Centre.\\
- {\em Swift} -- During the {\em INTEGRAL} monitoring of 3C~279, the source was observed nearly
simultaneously by {\em Swift}-XRT in photon counting mode on 2015-06-15 14:27 UTC (obs id: 00035019176),
which allows for a detailed spectrum in the energy range 0.6 -- 6.0 keV. In agreement with the very timely
analysis by \cite{pittori15a}, we find the spectrum to be affected by pile-up due to the bright flare and the hard
spectrum. After correcting for pile-up the spectral index is $\Gamma$=-1.37.\\
The same pointing by {\em Swift} led to observations with the UV-Optical Telescope (UVOT) with the $U$
filter.\\
- {\em SMARTS} -- During the {\em INTEGRAL} monitoring the blazar 3C~279 was also targeted by the Moderate
Aperture Research Telescope System (SMARTS) run by Yale University. Observations cover the optical to near-IR
bands $B$, $V$, $R$, $J$, and $K$.\\
- {\em FERMI} -- Also the Large Area Telescope (LAT) on board the {\em Fermi} gamma-ray mission captured the bright
flare of the source. A prompt analysis of this detection \cite{paliya15} allowed for a precise spectrum and lightcurve
of 3C~279.\\

\section{Spectral Energy Distribution}
\begin{figure}
\includegraphics[width=1.0\textwidth]{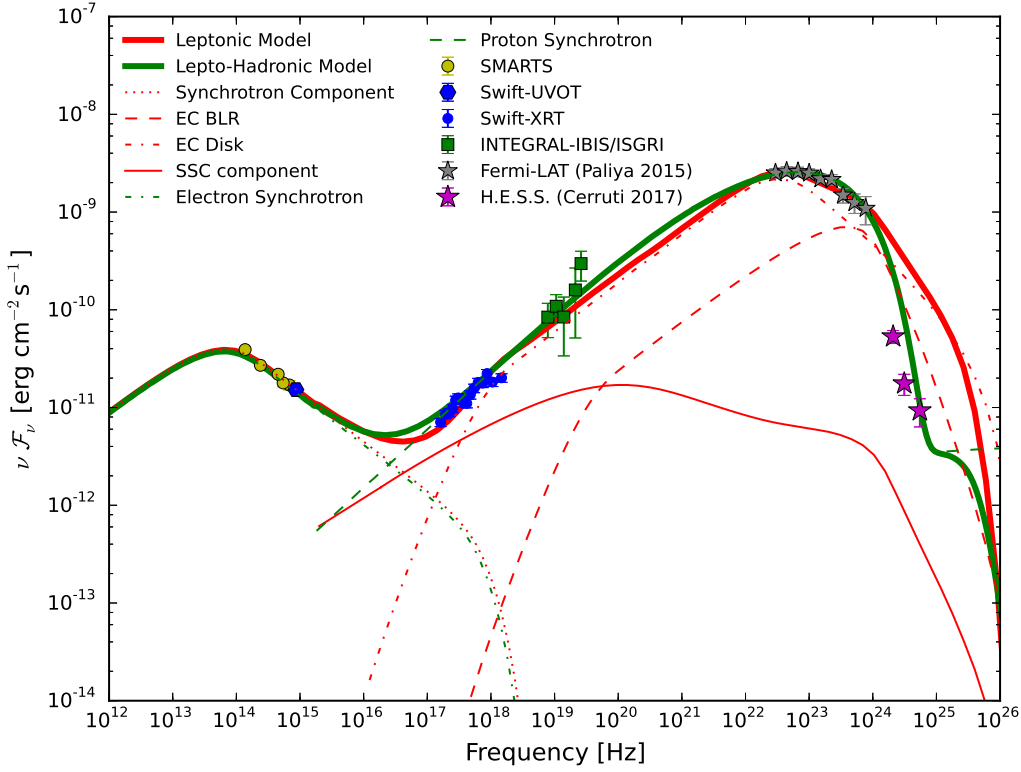}
\caption{Quasi-simultaneous SED of 3C~279, along with the leptonic (red solid) model and the lepto-hadronic
(green solid) model, as described in the text. The different emission components for both models as shown too (see
legend). The data from the H.E.S.S. collaboration were not included in the modeling. However, they line up surprisingly well
with the lepto-hadronic model.}
\label{fig:sed}
\end{figure}

The spectral energy distribution (SED) of the sources displays the two broad non-thermal radiation components 
characteristic of blazars (see Figure~\ref{fig:sed}). The low-energy component is  due to synchrotron emission
by relativistic electrons (possibly also positrons) in the jet, while the high-energy component can be either due to Compton scattering
by the same relativistic electrons (leptonic processes) or due to proton synchrotron radiation and synchrotron
emission from secondaries produced in photo-pion interactions (hadronic processes).\\
To model this boradband SED, we adopt the two  approaches: a leptonic and
a hadronic model. We use their time-independent homogeneous  one-zone jet radiation transfer \cite{boettcher13},
building upon an earlier work \cite{BC02}.

\section{Discussion and Conclusions}
The SED can be equally well modeled by a leptonic and by a lepto-hadronic model (see Figure~\ref{fig:sed}).
However, both models challenge the physical conditions in the jet.\\
The leptonic model indicates that the emission region is dominated
by kinetic energy by a factor of $\sim$~10 compared to the energy carried in the magnetic field. This may be difficult
to realize in a jet in which the magnetic field is the primary source of jet power because one would expect any mechanism
converting magnetic-field to particle kinetic energy to cease once equipartition is reached. In the leptonic scenario 
the emission region is inferred to be well within the broad line reagion (BLR), thus VHE photons would not escape this
region unattenuated \cite{boettcher16}. Therefore, the leptonic model predicts that no VHE photons should be
detected in this flare, if not originating at a different location.\\
On the other hand, our lepto-hadronic SED model allows us to choose parameters close to
equipartition between the magnetic field and the relativistic proton population. An important issue for this model may
be the extreme jet  power, of the order of the Eddington luminosity of the central black hole in 3C~279.\\
Because 3C~279 is one of currently only 5 FSRQs detected at VHE, the source is
subject to a ToO program by H.E.S.S. triggered on the basis of publicly available blazar observations.
Here we adapt the data of the 3C~279-flare presented in a very recent analysis by the H.E.S.S. collaboration
\cite{cerruti16}. The data are shown in Figure~\ref{fig:sed}. These data line up surprisingly well with our lepto-hadronic model.

\end{document}